\documentclass[twocolumn,aps,prc,superscriptaddress,showpacs,floatfix,longbibliography,nofootinbib]{revtex4-1}
\usepackage{url}
\usepackage{cancel}
\usepackage[colorlinks,linkcolor=blue,citecolor=blue,filecolor=black,urlcolor=blue]{hyperref}
\usepackage{epsfig,graphics}
\usepackage{graphicx}
\usepackage{dcolumn}
\usepackage{bm}
\usepackage[usenames]{color}
\usepackage{amssymb}
\usepackage{amsmath}
\usepackage{multirow}
\usepackage{float}
\usepackage{harpoon}
\usepackage{MnSymbol}
\usepackage{appendix}
\usepackage{color}
\usepackage{hyperref}
\usepackage{cleveref}
\usepackage{dutchcal}
\usepackage[normalem]{ulem}

\begin{document}

\title{Angular momentum conservation and pion production in intermediate-energy heavy-ion collisions}
\author{Hao-Nan Liu}
\affiliation{School of Physics Science and Engineering, Tongji University, Shanghai 200092, China}
\author{Rong-Jun Liu}\email[Rong-Jun Liu and Hao-Nan Liu contributed equally to this work.]{}
\affiliation{Shanghai Institute of Applied Physics, Chinese Academy of Sciences, Shanghai 201800, China}
\affiliation{University of Chinese Academy of Sciences, Beijing 100049, China}
\author{Jun Xu}\email[Correspond to\ ]{junxu@tongji.edu.cn}
\affiliation{School of Physics Science and Engineering, Tongji University, Shanghai 200092, China}
\affiliation{Southern Center for Nuclear-Science Theory, Institute of Modern Physics,
Chinese Academy of Sciences, Guangdong 516008, China}
\begin{abstract}
We have studied the effect of rigorous angular momentum conservation (AMC) in elastic, inelastic, and decay channels on pion production in intermediate-energy heavy-ion collisions based on the framework of an isospin-dependent Boltzmann-Uehling-Uhlenbeck (IBUU) transport model. We found that the constraint of AMC suppresses the absorption of both $\Delta$ resonances and pions, thus considerably enhances pion production and meanwhile reduces the charged pion yield ratio. The AMC effect on the charged pion yield ratio can not be simply compensated by a density-dependent in-medium $\Delta$ production cross section. Therefore, incorporating the constraint of AMC is important in obtaining the correct pion multiplicity and charged pion yield ratio by transport simulations, relevant for the extraction of the nuclear symmetry energy at high densities.
\end{abstract}
\maketitle


Understanding the equation of state of nuclear matter is one of the fundamental goals of nuclear physics, while the most uncertain part is the nuclear symmetry energy $E_{sym}$ which describes the energy difference between neutron-rich matter and isospin symmetric matter. The $E_{sym}$ at high densities has various ramifications in nuclear astrophysics~\cite{Steiner:2004fi,Lattimer:2006xb,Oertel:2016bki} and determines the neutron-richness of the high-density phase in intermediate-energy heavy-ion collisions~\cite{Baran:2004ih,Li:2008gp}. As first proposed in Refs.~\cite{Li:2002qx,Li:2002yda}, the charged pion yield ratio $\pi^-/\pi^+$ is strongly correlated with the isospin asymmetry in the high-density phase in intermediate-energy heavy-ion collisions, so it is an important experimental probe of the $E_{sym}$ at suprasaturation densities. While the extraction of the high-density $E_{sym}$ surfers from the model dependence of transport simulations~\cite{Xiao:2008vm,Feng:2009am,Xie:2013np}, we are now able to reduce the uncertainty of $\pi^-/\pi^+$ yield ratio to $1.6\%$ with reasonable physics input~\cite{TMEP:2023ifw}, thanks to the efforts by the transport model evaluation project. However, in order to accurately extract the $E_{sym}$ from the experimental data, cares must be taken to include complete physics related to pion production, such as the threshold effect~\cite{Song:2015hua}, the pion in-medium effect~\cite{Zhang:2017mps,Hong:2013yva,Xu:2013aza,Xu:2009fj}, the clustering effect~\cite{Ikeno:2016xpr}, etc.

Since pions are second-order observables compared to those of nucleons in intermediate-energy heavy-ion collisions, their production could be affected by detailed energy or angular momentum conservation. Effects of the conservation law on the dynamics of intermediate-energy heavy-ion collisions were first explored in Ref.~\cite{Gale:1990zz}. The study in Ref.~\cite{Cozma:2014yna} has investigated how various options of the energy conservation may affect pion production. The effect of the AMC in nucleon-nucleon elastic collisions on overall dynamics was revisited in Ref.~\cite{Liu:2023pgc}, where the AMC in inelastic collisions has not been treated consistently. Later on, the interplay of the rigorous AMC and the spin dynamics in intermediate-energy heavy-ion collisions was investigated in Ref.~\cite{Liu:2023nkm}.

In the present study, we will incorporate rigorous AMC into channels relevant to pion production, i.e., $N+N \leftrightarrow N+\Delta$ and $\Delta \leftrightarrow N+\pi$, in the IBUU transport model, and study the effect on the final pion multiplicity and $\pi^-/\pi^+$ yield ratio. We will show that the AMC suppresses the $\Delta$ absorption channel $N+\Delta \rightarrow N+N$ and the pion absorption channel $N+\pi \rightarrow \Delta$, so the net production of pions is enhanced while the $\pi^-/\pi^+$ yield ratio is reduced. Therefore, neglecting the constraint of AMC may influence the accurate extraction of the $E_{sym}$ at high densities.


The IBUU transport model solves numerically the following IBUU equation using the test-particle method~\cite{Bertsch:1988ik}
\begin{equation}
  \label{eq:boltz}
  \frac{\partial f_\tau(\vec{r},\vec{p})}{\partial t}
  +\frac{\vec{p}}{\sqrt{m^2+p^2}}\cdot
  \frac{\partial f_\tau(\vec{r},\vec{p})}{\partial \vec{r}}-\frac{\partial U_\tau}{\partial \vec{r}}\cdot\frac{\partial f_\tau(\vec{r},\vec{p})}{\partial \vec{p}}
  =I_c.
\end{equation}
The left-hand side of the above equation describes how the one-body phase-space distribution function $f_\tau(\vec{r},\vec{p})$ evolves under the mean-field potential $U_\tau$, with $\tau=n,p$ being the isospin index. For the purpose of illustration on the AMC effect, we use the following momentum-independent mean-field potential
\begin{eqnarray}\label{u}
U_{n/p}(\rho,\delta)=\alpha \left(\frac{\rho}{\rho_{0}}\right)+\beta\left(\frac{\rho}{\rho_{0}}\right)^{\gamma} \pm 2 E_{sym}^{pot}  \left(\frac{\rho}{\rho_{0}}\right)^{\gamma_{sym}} \delta, 
\end{eqnarray}
where $\rho_0=0.16$ fm$^{-3}$ is the saturation density, $\rho=\rho_n+\rho_p$ is the nucleon number density, $\delta=(\rho_n-\rho_p)/\rho$ is the isospin asymmetry, and the `$+$' (`$-$') sign is used for $U_n$ ($U_p$). We set $\alpha=209.2$ MeV, $\beta=156.4$ MeV, $\gamma=1.35$, $E_{sym}^{pot}=18$ MeV, and $\gamma_{sym}=1.1$ to reproduce empirical properties of isospin asymmetric nuclear matter. The collision integral $I_c$ on the right-hand side includes both elastic and inelastic collisions. A constant and isotropic cross section of 40 mb is used for elastic baryon-baryon scatterings. The channels $N+N \leftrightarrow N+\Delta$ and  $\Delta \leftrightarrow N+\pi$ related to pion production are also incorporated, with detailed cross sections and decay widths set as the same as in Ref.~\cite{TMEP:2023ifw}, where the detailed balance condition for the forward and backward reactions is satisfied.

To properly incorporate the constraint of AMC in above inelastic channels, especially for $N+\pi \rightarrow \Delta$, the spin degree of freedom must be included in IBUU. Instead of performing a comprehensive simulation of spin dynamics as in Ref.~\cite{Liu:2023nkm}, in the present study we neglect the spin-orbit potential, and give randomized spins for nucleons in the initialization and after each scattering. While pions have zero spin, the spin of $\Delta$ resonances should be treated with care. The spin state of a spin-$3/2$ $\Delta$ resonance projected on the $z$ direction can be represented by
$$
|\psi_\Delta\rangle = \begin{pmatrix}
c_{3/2} \\
c_{1/2} \\
c_{-1/2} \\
c_{-3/2}
\end{pmatrix},
$$
with the complex probability coefficients satisfying the normalization condition
\begin{equation}\label{nor}
|c_{3/2}|^2 + |c_{1/2}|^2 + |c_{-1/2}|^2 + |c_{-3/2}|^2 = 1.
\end{equation}
The spin operators for the spin-$3/2$ $\Delta$ resonance are given by the $4\times 4$ matrices, i.e.,
$$
\Sigma_x = \frac{1}{2} \begin{pmatrix}
0 & \sqrt{3} & 0 & 0 \\
\sqrt{3} & 0 & 2 & 0 \\
0 & 2 & 0 & \sqrt{3} \\
0 & 0 & \sqrt{3} & 0
\end{pmatrix},
$$
$$
\Sigma_y = \frac{1}{2} \begin{pmatrix}
0 & -i\sqrt{3} & 0 & 0 \\
i\sqrt{3} & 0 & -2i & 0 \\
0 & 2i & 0 & -i\sqrt{3} \\
0 & 0 & i\sqrt{3} & 0
\end{pmatrix},
$$
$$
\Sigma_z = \begin{pmatrix}
3/2 & 0 & 0 & 0 \\
0 & 1/2 & 0 & 0 \\
0 & 0 & -1/2 & 0 \\
0 & 0 & 0 & -3/2
\end{pmatrix}.
$$
For a given spin state $|\psi_\Delta\rangle$, each component of the spin expectation vector $\vec{s}_\Delta = (s_{\Delta x}, s_{\Delta y}, s_{\Delta z})$ can be calculated from
$$
s_{\Delta i} = \langle \psi_\Delta | \Sigma_i | \psi_\Delta \rangle, \quad (i = x, y, z).
$$
When a $\Delta$ resonance is produced through the $N+N \rightarrow N+ \Delta$ or $N+\pi \rightarrow \Delta$ process in IBUU, its spin $\vec{s}_\Delta$ is randomly sampled, i.e., both real and imaginary parts of $c_{3/2}$, $c_{1/2}$, $c_{-1/2}$, and $c_{-3/2}$ are sampled according to a normal distribution under the constraint of the normalization condition [Eq.~(\ref{nor})]. We note that in this way $|\vec{s}_\Delta|$ is not uniformly distributed within $[0,3/2]$.

For the AMC in the $N+N \leftrightarrow N+\Delta$ channels, we generalize the method in Ref.~\cite{Liu:2023nkm} to inelastic $2 \leftrightarrow 2$ processes. The AMC condition can be expressed as
\begin{align*}
\vec{R}\times \vec{P}+\vec{r}\times \vec{p}+\vec{S}_{\text{tot}}=\vec{R}^{'}\times \vec{P}^{'}+\vec{r}^{'}\times \vec{p}^{'}+\vec{S}_{\text{tot}}^{'},
\end{align*}
where
\begin{eqnarray}
&&\vec{R}=(\vec{r}_{1}+\vec{r}_{2})/2, \vec{r}=(\vec{r}_{1}-\vec{r}_{2})/2, \notag\\
&&\vec{P}=\vec{p}_1+\vec{p}_2, \vec{p}=\vec{p}_{1}-\vec{p}_{2}, \notag\\
&&\vec{S}_{\text{tot}}=\vec{s}_1+\vec{s}_2 \notag
\end{eqnarray}
are the centroid coordinate, relative coordinate, centroid momentum, relative momentum, and total spin, respectively, in the initial state, with $\vec{r}_{1(2)}$, $\vec{p}_{1(2)}$, and $\vec{s}_{1(2)}$ being the coordinate, momentum, and spin expectation vector of particle $1(2)$, and quantities with `$\prime$' represent those in the final state. By setting $\vec{R}=\vec{R}^{'}$, applying momentum conservation $\vec{P}=\vec{P}^{'}$, and randomly sampling $\vec{s}^{'}_1$ and $\vec{s}^{'}_2$, the task of AMC is to find proper $\vec{r}^{'}$ and $\vec{p}^{'}$ which lead to a suitable final-state relative angular momentum $\vec{L}^{'} = \vec{r}^{'} \times \vec{p}^{'}$. The idea is to select properly the direction of the center-of-mass (C.M.) momentum $\vec{p}_1^{*'}$ so that the relative momentum $\vec{p}^{'}$ is perpendicular to $\vec{L}^{'}$. This is identical to requiring
\begin{equation}
\vec{p}^{'} \cdot \vec{L}^{'}={\vec{p}_1^{*'}}\cdot \left( 2\vec{L}^{'}+\frac{2\gamma ^2}{\gamma +1}\vec{\beta }\cdot \vec{L}^{'}\vec{\beta} \right) +\gamma\Delta e^{*}\vec{\beta }\cdot \vec{L}^{'}=0.\\
\end{equation}
In the above, $\vec{\beta}$ is the velocity of the C.M. frame with respective to the lab frame, $\gamma = 1/\sqrt{1-\beta^2}$ is the Lorentz factor, and $\Delta e^*=\sqrt{{{p}_{1}^{*'}}^2+{m}_{1}^{2}}-\sqrt{{{p}_{2}^{*'}}^2+{m}_{2}^{2}}$ is the energy difference between two particles after scatterings in the C.M. frame, which is zero for the $N+\Delta \rightarrow N+N$ process but nonzero for the $N+N \rightarrow N+\Delta$ process.

Defining $\vec{q}_1=2\vec{L}^{'}+\frac{2\gamma ^2}{\gamma +1}\vec{\beta }\cdot \vec{L}^{'}\vec{\beta}$ and $b=\gamma\Delta e^{*}\vec{\beta }\cdot \vec{L}^{'}$, and for a given scattering polar angle $\theta$ in the C.M. frame, $\vec{p}_1^{*'}$ must simultaneously satisfy both constraints of $\vec{p}_1^{*'}\cdot \vec{q}_1 = -b$ and $\vec{p}_1^{*'}\cdot \vec{p}_1^{*}={p_1^{*'}} p_1^{*}\cos \theta$. The two constraints can be combined into the following condition
$$
\vec{p}_1^{*'} \cdot \left[ (p_1^{*'} p_1^{*}\cos \theta) \vec{q}_1 + b \vec{p}_1^{*} \right] = 0,
$$
showing that $\vec{p}_1^{*'}$ must be perpendicular to
\begin{equation}
\vec{q}_2 = (p_1^{*'} p_1^{*}\cos \theta) \vec{q}_1 + b \vec{p}_1^{*}.
\end{equation}
To satisfy the above constraint, we select the direction of $\vec{p}_1^{*'}$ by setting
\begin{equation}
{\vec{p}_1^{*'}}={p_1^{*'}}\left( \cos \varphi \vec{\hat{e}}_1+\sin \varphi \vec{\hat{e}}_2 \right), \label{p1s_inelastic}
\end{equation}
with $\vec{\hat{e}}_1$ and $\vec{\hat{e}}_2$ being the unit vectors in the directions of $\vec{q}_2\times \vec{p}_1^{*}$ and $\vec{q}_2\times \vec{p}_1^{*}\times \vec{q}_2$, respectively, and the angle $\varphi$ gives the direction of $\vec{p}_1^{*'}$ in the plane formed by $\vec{\hat{e}}_1$ and $\vec{\hat{e}}_2$. Substituting Eq.~(\ref{p1s_inelastic}) into $\vec{p}_1^{*'}\cdot \vec{p}_1^{*}={p_1^{*'}} p_1^{*}\cos \theta$ leads to
\begin{equation}
\sin \varphi =\frac{p_1^{*} \cos \theta}{\vec{\hat{e}}_2\cdot \vec{p}_1^{*}}.
\end{equation}
No solution will be obtained in the case of $\left| \cos \theta \right| > \left| \vec{\hat{e}}_2\cdot \vec{p}_1^*/p_1^* \right|$, largely due to the randomized spin in the final state of the scattering, so in real simulations $\cos\theta$ is sampled in a smaller range that satisfies $|\sin \varphi| \le 1$.

The above method puts the direction of $\vec{r}^{'}\times \vec{p}^{'}$ to be the same as $\vec{L}^{'}$. To make them have the same magnitude, the relative coordinates $\vec{r}^{'}$ of the colliding particles must satisfy
\begin{equation}
\vec{r}^{'}= A \vec{\hat{e}}_{\vec{p}^{'}}+\frac{L^{'}}{p^{'}}\vec{\hat{e}}_{\vec{p}^{'}\times \vec{L}^{'}} ,
\end{equation}
where $\vec{\hat{e}}_{\vec{p}^{'}}$ and $\vec{\hat{e}}_{\vec{p}^{'}\times \vec{L}^{'}}$ are the unit vectors in the direction of $\vec{p}^{'}$ and $\vec{p}^{'}\times \vec{L}^{'}$, respectively, and $A$ is set to be $\vec{r}\cdot \vec{\hat{e}}_{\vec{p}^{'}}$ in order to minimize $|\vec{r}^{'}-\vec{r}|$.

\begin{figure}[ht]
\includegraphics[width=0.8\linewidth]{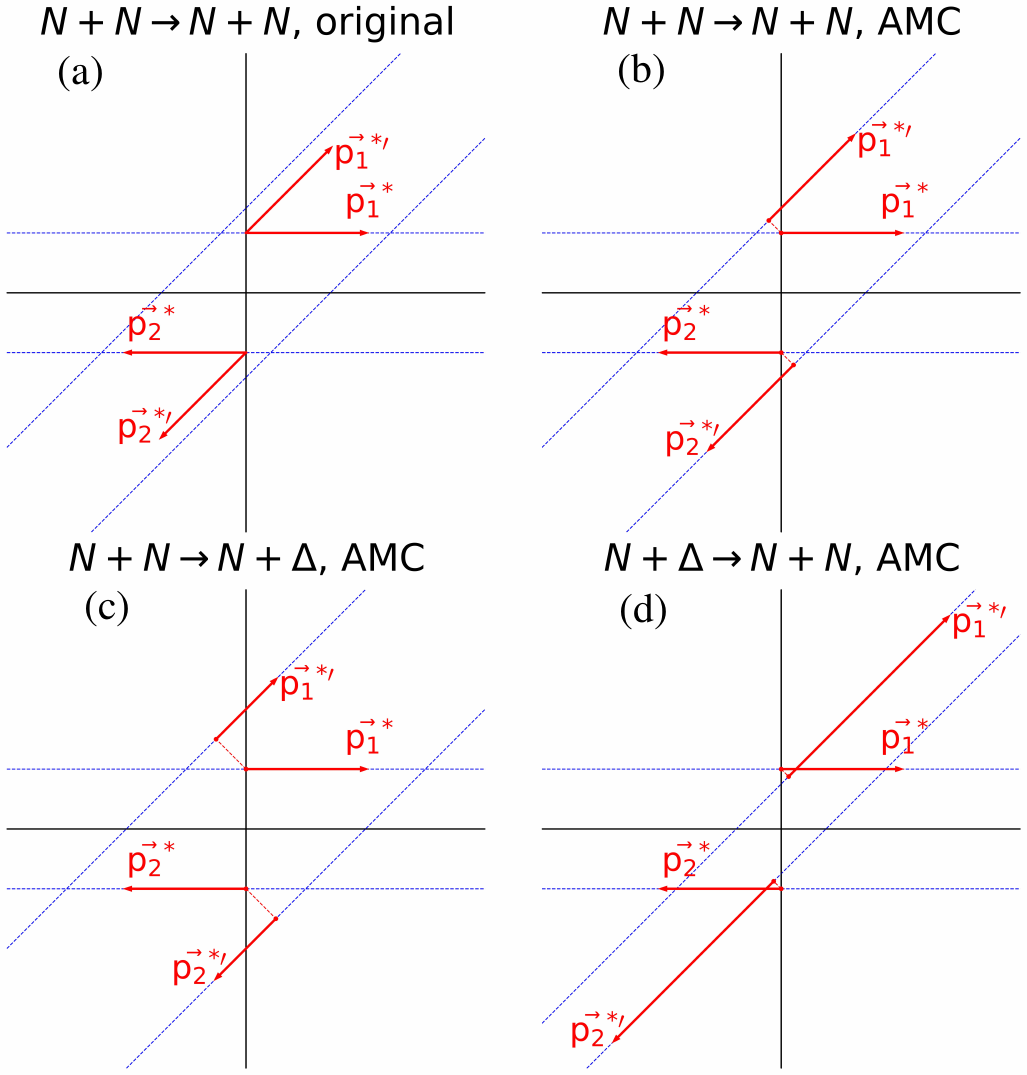}
\caption{\label{fig1} Cartoons for the initial-state and final-state momenta in the C.M. frame of $N+N\rightarrow N+N$ scatterings without (a) and with AMC (b), $N+N\rightarrow N+\Delta$ scatterings with AMC (c), and $N+\Delta\rightarrow N+N$ scatterings with AMC (d).}
\end{figure}

Figure~\ref{fig1} displays the cartoons for the elastic and inleastic $2 \leftrightarrow 2$ processes in the C.M. frame. Figure~\ref{fig1} (a) represents the original treatment of elastic nucleon-nucleon (NN) scatterings, which generally does not conserve angular momentum, unless the impact parameter remains unchanged after NN scatterings, and this can be achieved by moving the coordinates of final-state nucleons by a shortest distance perpendicular to the final momentum, as shown in Fig.~\ref{fig1} (b). For the $N+N \rightarrow N+\Delta$ process, since the magnitude of the final-state momentum is generally smaller due to the large mass of $\Delta$ resonance, the final-state particles are moved by a longer distance, as shown in Fig.~\ref{fig1} (c). This effect enhances the separation between nucleons and $\Delta$ resonances, and suppresses the subsequent $\Delta$ absorption process to be detailed later. For the $N+\Delta \rightarrow N+N$ process, the final-state nucleons may have larger momenta and smaller separation distance, but this effect could be partially cancelled by the separation of final-state nucleons in elastic NN scatterings as shown in Fig.~\ref{fig1} (b). If the test particle of baryons is not a point particle but has a final size, e.g., in the lattice Hamiltonian framework~\cite{Lenk:1989zz}, the jump of the final-state coordinate then represents that of the centroid coordinate in the shape function of the test particle, and the framework is still valid.

We now derive the angular distribution amplitude from the decay of a $\Delta$ spin state $|\psi_\Delta\rangle$ into a nucleon of spin $s_N$ and a pion. The amplitude $A_{m_s}(\theta, \phi)$ for finding the nucleon in a final spin state $m_s$ is given by the S-matrix element~\cite{Jacob:1959at}:
\begin{equation}
A_{m_s}\left( \theta ,\phi \right) = \left< \psi _{\text{final}} \right|\hat{S}\left| \psi _{\Delta} \right> = \sum_{M=-3/2}^{3/2}{c_M\left< \theta ,\phi ;s_N,m_s \right|\hat{S}\left| J,M \right>}, \notag\\
\end{equation}
where  $\left| \psi _{\text{final}} \right> =\left| \theta ,\phi \right> \otimes \left| s_N,m_s \right>$ represents the final state of the decay, with $\theta$ and $\phi$ being the polar and azimuthal angles of nucleon momentum in the $\Delta$ rest frame. By inserting a complete set of angular momentum states $\sum_{L,m_L,m}{\left| L,m_L;s_N,m \right>}\left< L,m_L;s_N,m \right|$ and using the relation $\left< \theta ,\phi ;s_N,m_s\mid L,m_L;s_N,m \right> = \left< \theta ,\phi \mid L,m_L \right> \left< s_N,m_s \mid s_N,m \right> =\delta _{m_s,m}Y_{L,m_L}\left( \theta ,\phi \right)$, the above amplitude can be rewritten as
\begin{eqnarray}
&&~~~A_{m_s}\left( \theta ,\phi \right) \notag\\
&&= \sum_{M}{c_M\sum_{L,m_L}{\left< L,m_L;s_N,m_s \right|\hat{S}\left| J,M \right>}}Y_{L,m_L}\left( \theta ,\phi \right) \notag\\
&&= \sum_{M}{c_M\sum_{L,m_L}{A_{L,s_N,J}\left< L,m_L;s_N,m_s \mid J,M \right>}}Y_{L,m_L}\left( \theta ,\phi \right). \notag
\end{eqnarray}
To get the last line in the above equation, we have used the Wigner-Eckart theorem~\cite{wigner2012group} so that the matrix element is proportional to the Clebsch-Gordan coefficient $\left< L,m_L;s_N,m_s \mid J,M \right>$. $A_{L,s_N,J}$ is the reduced matrix element, which is essentially a constant given $J=3/2$, $s_N=1/2$, and $L=1$ as a result of the parity conservation in the strong decay, and can be absorbed in the normalization. The final expression of the amplitude is
$$
A_{m_s}(\theta, \phi) = \sum_{M=-3/2}^{3/2} c_M \sum_{m_L=-1}^{1} \left\langle 1, m_L; \frac{1}{2}, m_s \bigg| \frac{3}{2}, M \right\rangle Y_{1, m_L}(\theta, \phi),
$$
with $\left\langle 1, m_L; \frac{1}{2}, m_s \bigg| \frac{3}{2}, M \right\rangle$ being the Clebsch-Gordan coefficients and $Y_{1,m_L}(\theta, \phi)$ being the spherical harmonics. The differential decay rate is proportional to the sum of the squared amplitudes
\begin{equation}\label{Gamma}
\frac{d\Gamma}{d\Omega} \propto \left| A_{1/2} \right|^2+\left| A_{-1/2} \right|^2.
\end{equation}

The AMC in the $\Delta \rightarrow N+\pi$ process requires
\begin{equation}
\vec{R}_\Delta \times \vec{P}_\Delta +\vec{s}_\Delta=\vec{R}^{'}\times \vec{P}^{'}+\vec{r}^{'}\times \vec{p}^{'}+\vec{s}^{'}_N,
\end{equation}
where $\vec{R}_\Delta$, $\vec{P}_\Delta$, and $\vec{s}_\Delta$ are the coordinate, momentum, and spin expectation vector of the initial $\Delta$, respectively, $\vec{R}^{'}$ and $\vec{P}^{'}$ are the centroid coordinate and momentum in the final state consisting of a nucleon and a pion, $\vec{r}^{'}$ and $\vec{p}^{'}$ are the corresponding relative coordinate and momentum in the final state, and $\vec{s}'_N$ is the nucleon spin expectation vector. By requiring $\vec{R}_\Delta = \vec{R}^{'}$ and using the momentum conservation condition $\vec{P}_\Delta = \vec{P}^{'}$, the above AMC condition is reduced to
\begin{equation}\label{decayamc}
\vec{L}^{'} = \vec{r}^{'}\times \vec{p}^{'} = \vec{s}_\Delta - \vec{s}^{'}_N.
\end{equation}
Next, we first sample the nucleon momentum $\vec{p}^{*'}$ in the rest frame of $\Delta$ resonance according to Eq.~(\ref{Gamma}), which gives $\vec{p}^{'}$ in Eq.~(\ref{decayamc}) after proper Lorentz transformation. To satisfy AMC, the orbital angular momentum $\vec{L}^{'}$ must be perpendicular to $\vec{p}^{'}$, i.e., $\vec{L}^{'} \cdot \vec{p}^{'}=0$, which leads to
\begin{equation}
(\vec{s}_\Delta - \vec{s}^{'}_N) \cdot \vec{p}^{'}=0,
\end{equation}
and this gives the polar angle $\theta_{sp}^N$ between $\vec{s}^{'}_N$ and $\vec{p}^{'}$
\begin{equation}
\cos \theta_{s p}^N = \frac{\vec{s}_\Delta \cdot \vec{p}^{'} }{|\vec{s}_N^{'}| p^{'}},
\end{equation}
with $|\vec{s}_N^{'}|$ being $1/2$. If the resulting $|\cos \theta_{s p}^N|$ is larger than 1, $\vec{p}^{*'}$ should be resampled in order to get a proper $\vec{p}^{'}$ until $|\cos \theta_{s p}^N| \le 1$ is satisfied. The corresponding azimuthal angle is uniformly sampled within $[0, 2\pi]$, so that the direction of $\vec{s}_N^{'}$ can be determined. Finally, the relative coordinate in the final state is given by
\begin{equation}
\vec{r}^{'} = \frac{1}{{p^{'}}^2} \vec{p}^{'} \times \vec{L}^{'}.
\end{equation}

In the inverse $N+\pi \rightarrow \Delta$ process, the total angular momentum $\vec{J}$ in the initial state contains the contributions from the orbital angular momentum and the nucleon spin. The AMC condition requires
\begin{equation}\label{LNpiD}
\vec{L}^{'} =\vec{R}_\Delta^{'}\times \vec{P}_\Delta^{'} = \vec{J} - \vec{s}_\Delta^{'},
\end{equation}
where $\vec{R}_\Delta^{'}$ and $\vec{P}_\Delta^{'}$ are the coordinate and momentum of $\Delta$, respectively, and $\vec{s}_\Delta^{'}$ is the $\Delta$ spin expectation vector. Here $\vec{P}_\Delta^{'}$ is known from momentum conservation, while $\vec{R}_\Delta^{'}$ and $\vec{s}_\Delta^{'}$ are to be determined. The magnitude of $\Delta$ spin $|\vec{s}_\Delta^{'}|$ is randomly sampled, while its direction is determined in the following way. Taking the dot product with $\vec{P}_\Delta^{'}$ on both sides of Eq.~(\ref{LNpiD}) leads to
\begin{equation}
\vec{J} \cdot \vec{P}_\Delta^{'}  = \vec{s}_\Delta^{'}  \cdot \vec{P}_\Delta^{'},
\end{equation}
which determines the polar angle $\theta_{s p}^\Delta$ between the $\Delta$ spin $\vec{s}_\Delta^{'} $ and momentum $\vec{P}_\Delta^{'}$
\begin{equation}\label{costhetasp}
\cos \theta_{s p}^\Delta = \frac{\vec{J} \cdot \vec{P}_\Delta^{'} }{|\vec{s}_\Delta^{'} | P_\Delta^{'} }.
\end{equation}
The corresponding azimuthal angle is uniformly sampled within $[0, 2\pi]$, so that the direction of $\vec{s}_\Delta^{'}$ can be determined. With $\vec{s}_\Delta^{'}$ determined and $\vec{L}^{'}$ obtained from Eq.~(\ref{LNpiD}), the coordinate $\vec{R}_\Delta^{'}$ must lie in the plane spanned by $\vec{P}_\Delta^{'}$ and $\vec{P}_\Delta^{'} \times \vec{L}^{'}$, and has the following general form
\begin{equation}
\vec{R}_\Delta^{'} = A \vec{\hat{e}}_{\vec{P}_\Delta^{'}} + \frac{L^{'}}{P_\Delta^{'}} \vec{\hat{e}}_{\vec{P}_\Delta^{'} \times \vec{L}^{'}},
\end{equation}
where $\vec{\hat{e}}_{\vec{P}_\Delta^{'}}$ and $\vec{\hat{e}}_{\vec{P}_\Delta^{'} \times \vec{L}^{'}}$ represent the unit vectors in the directions of $\vec{P}_\Delta^{'}$ and $\vec{P}_\Delta^{'} \times \vec{L}^{'}$, respectively. Here $A$ is set to be $\vec{R}_{CM} \cdot \vec{\hat{e}}_{\vec{P}_\Delta^{'}}$ in order to minimize the distance between the coordinate of the final-state $\Delta$ resonance and the C.M. position $\vec{R}_{CM}$ of the initial $\pi N$ system.

We note that adjusting $\vec{R}_\Delta^{'}$ may compensate the angular momentum contribution perpendicular to $\vec{P}_\Delta^{'}$, while the contribution along $\vec{P}_\Delta^{'}$ can only be compensated by the $\Delta$ spin, which can't be too large otherwise the $|\cos \theta_{s p}^\Delta|$ in Eq.~(\ref{costhetasp}) becomes larger than 1. In such case, the value of $|\vec{s}_\Delta^{'}|$ is sampled again, while the process still may not be allowed since $|\vec{s}_\Delta^{'}|$ has a maximum value of $3/2$. An obtained spin expectation vector $\vec{s}_\Delta^{'}$ corresponds to an infinite number of spin pure states $\lvert\psi_\Delta\rangle$, and a gradient descent method is employed to randomly find one such spin state $\lvert\psi_\Delta\rangle$. For the same $\vec{s}_\Delta^{'}$, different $\lvert\psi_\Delta\rangle$ states will result in different final-state angular distributions for the decay according to Eq.~(\ref{Gamma}). While the rotational symmetry of the angular distribution perpendicular to $\vec{s}_\Delta^{'}$ may not be satisfied for arbitrary $\lvert\psi_\Delta\rangle$ state, it is satisfied by averaging all possible $\lvert\psi_\Delta\rangle$ states.


\begin{figure}[ht]
\includegraphics[width=0.7\linewidth]{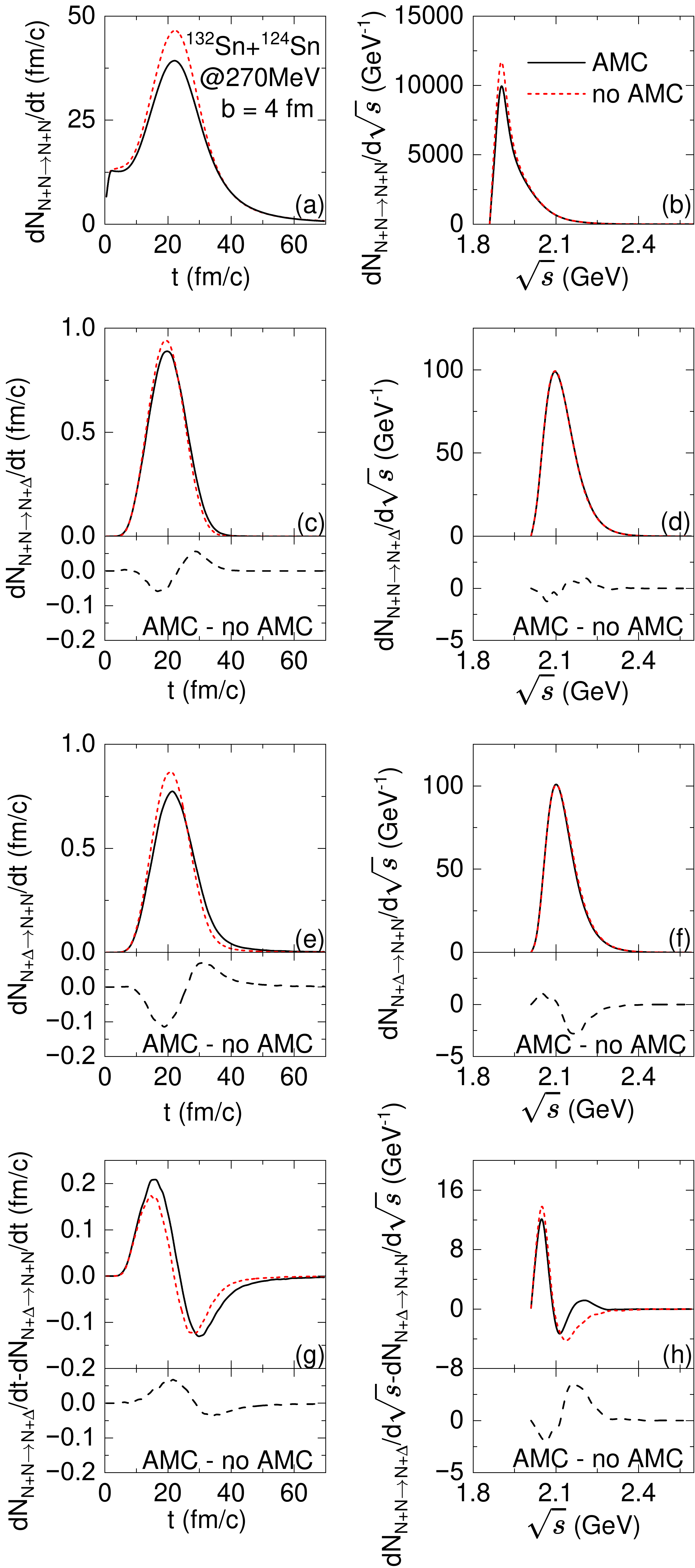}
\caption{\label{fig2} Comparison of the collision rates as a function of time (left) and C.M. energy (right) for $N+N\rightarrow N+N$ scatterings (first row), $N+N\rightarrow N+\Delta$ scatterings (second row), $N+\Delta\rightarrow N+N$ scatterings (third row), and net $\Delta$ production (fourth row) with and without the constraint of AMC in $^{132}$Sn+$^{124}$Sn collisions at the beam energy of 270 AMeV and an impact parameter $\text{b}=4$ fm.}
\end{figure}

\begin{figure}[ht]
\includegraphics[width=0.9\linewidth]{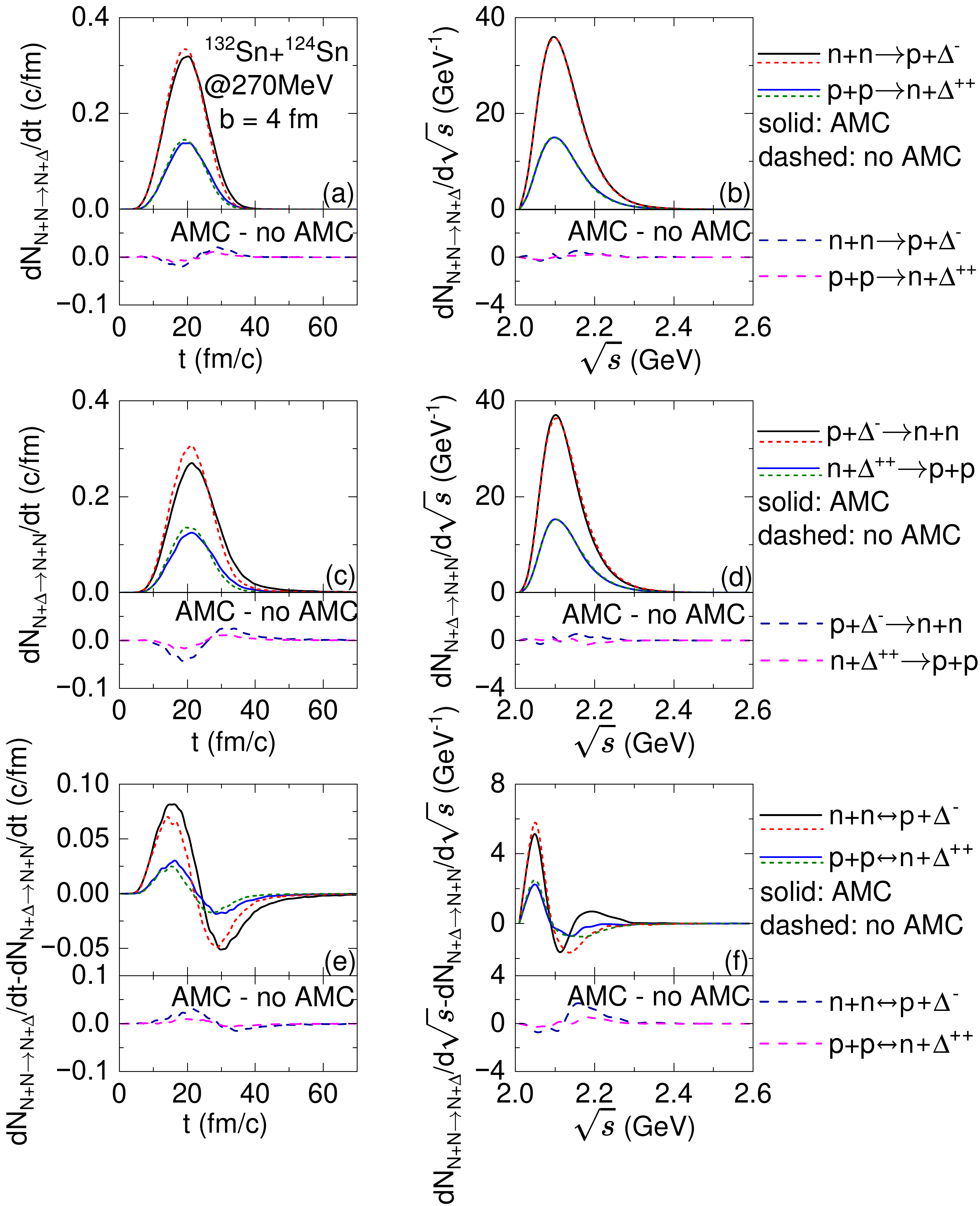}
\caption{\label{fig3} Comparison of the collision rates as a function of time (left) and C.M. energy (right) for $n+n\rightarrow p+\Delta^{++}$ and $p+p\rightarrow n+\Delta^{-}$ scatterings (first row), inverse processes (second row), and net $\Delta$ production (third row) with and without the constraint of AMC in $^{132}$Sn+$^{124}$Sn collisions at the beam energy of 270 AMeV and an impact parameter $\text{b}=4$ fm.}
\end{figure}

\begin{figure}[ht]
\includegraphics[width=0.7\linewidth]{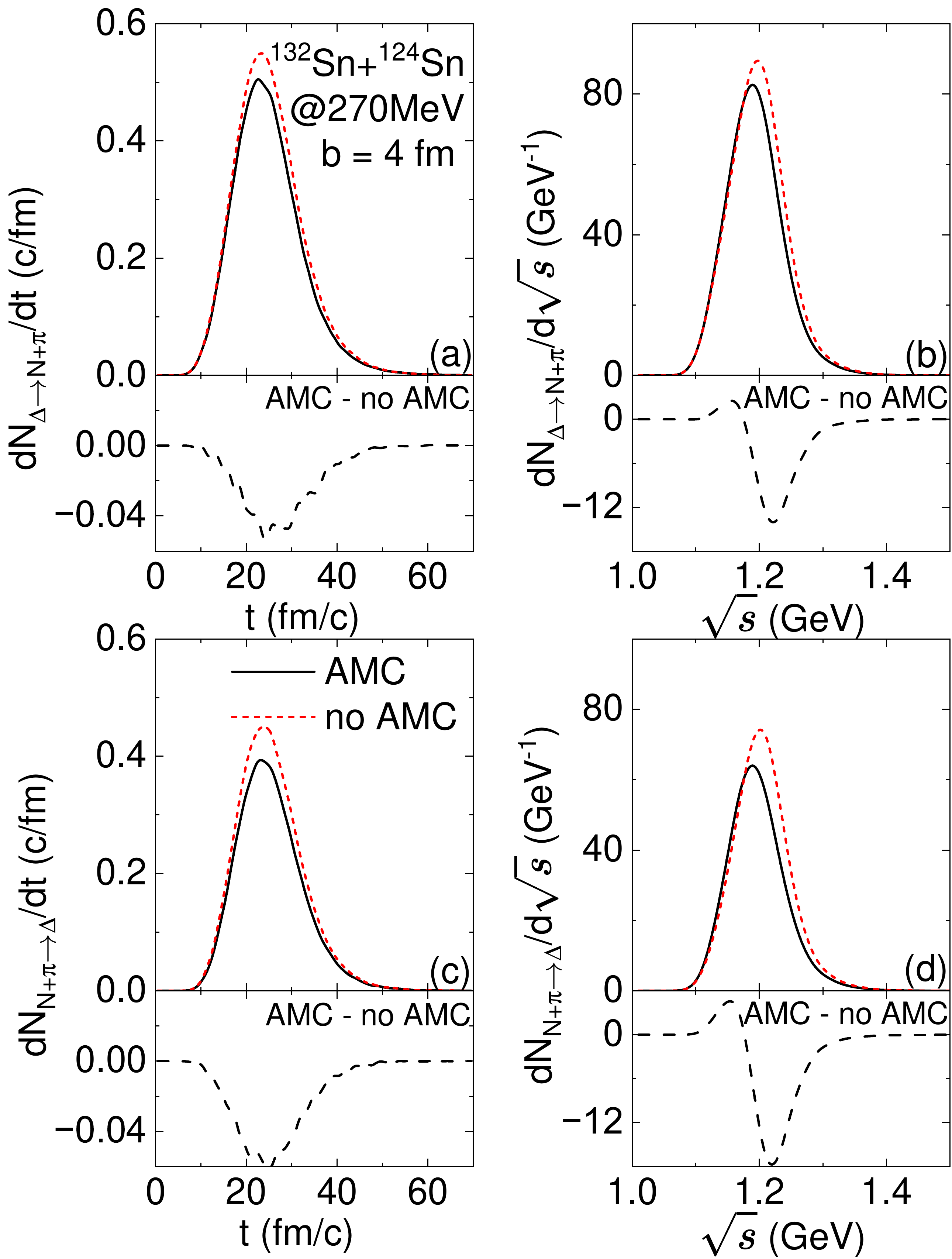}
\caption{\label{fig4} Comparison of the reaction rates as a function of time (left) and C.M. energy (right) for $\Delta \rightarrow N+\pi$ processes (first row) and $N+\pi \rightarrow \Delta$ processes (second row) with and without the constraint of AMC in $^{132}$Sn+$^{124}$Sn collisions at the beam energy of 270 AMeV and an impact parameter $\text{b}=4$ fm.}
\end{figure}

\begin{figure}[ht]
\includegraphics[width=0.5\linewidth]{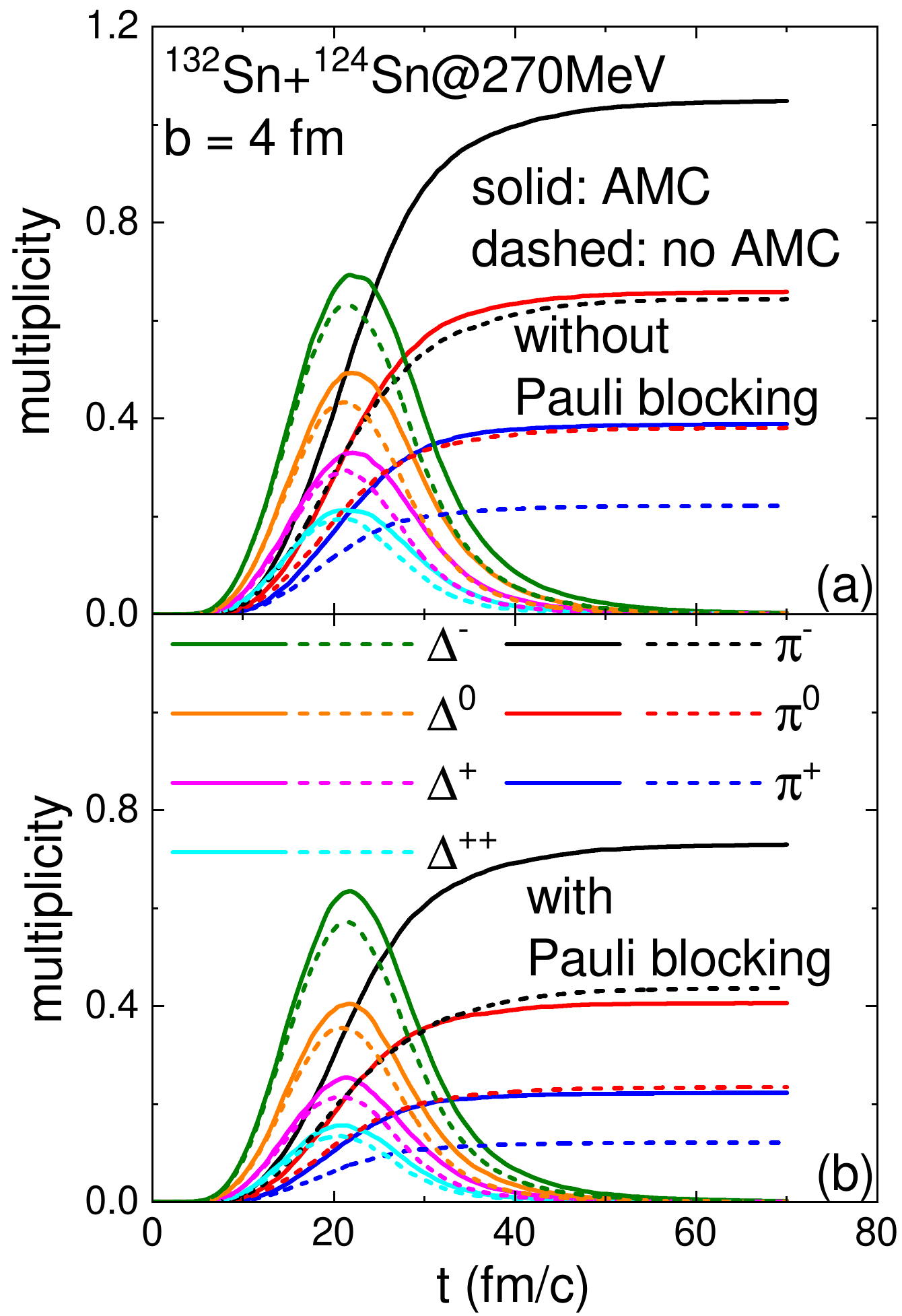}
\caption{\label{fig5} Production of pion-like particles with time evolution with and without the constraint of AMC in $^{132}$Sn+$^{124}$Sn collisions at the beam energy of 270 AMeV and an impact parameter $\text{b}=4$ fm, with the upper panel for the case without Pauli blocking and lower panel for the case with Pauli blocking.}
\end{figure}

\begin{figure}[ht]
\includegraphics[width=0.9\linewidth]{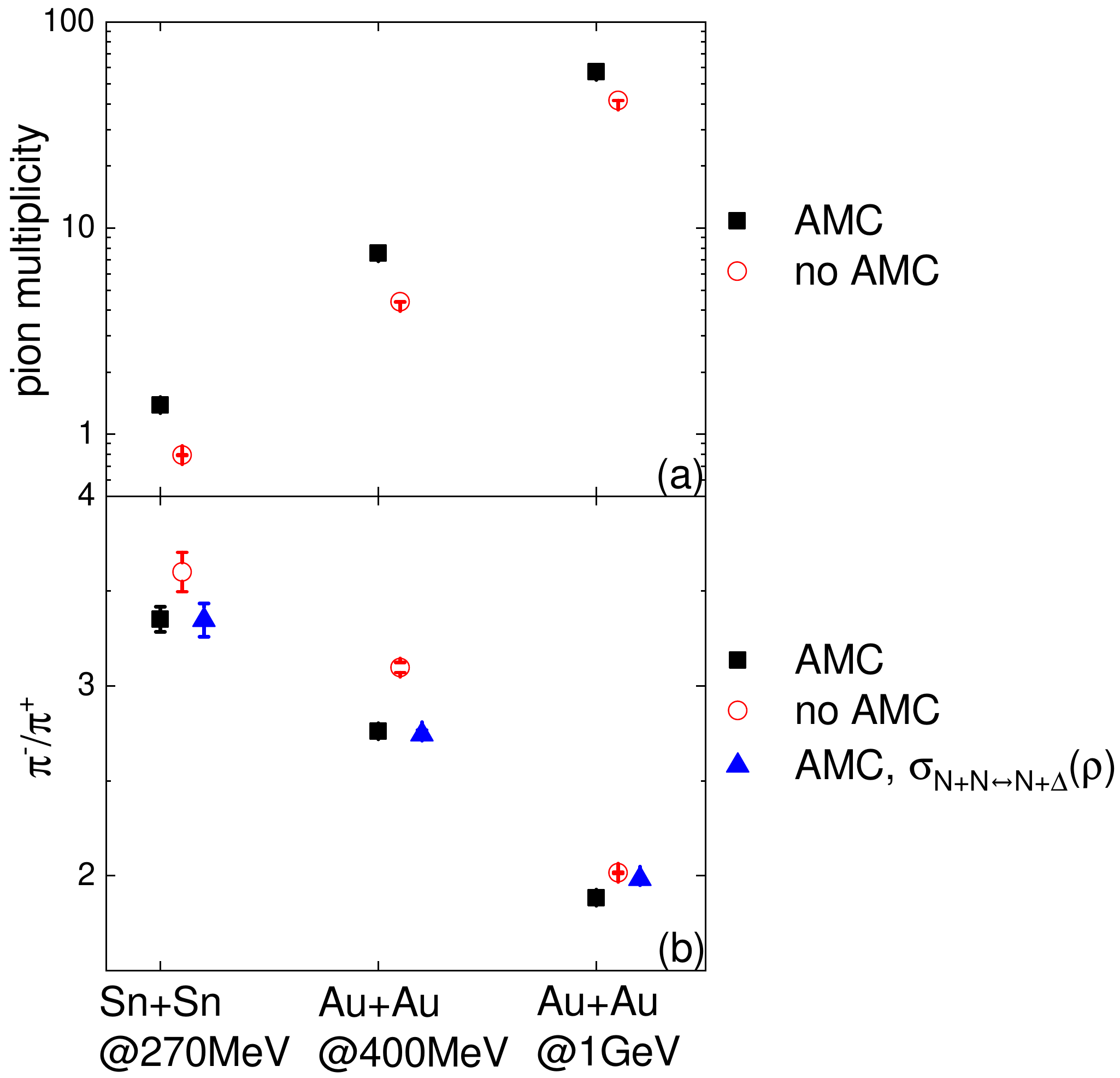}
\caption{\label{fig6} Upper: Total pion multiplicities in Sn+Sn and Au+Au collisions at different beam energies with and without the constraint of AMC; Lower: $\pi^-/\pi^+$ yield ratios in Sn+Sn and Au+Au collisions at different beam energies with and without the constraint of AMC, as well as those with in-medium cross sections for $N+N \leftrightarrow N+\Delta$ processes to reproduce the same pion multiplicities.}
\end{figure}

We perform IBUU simulations for $^{132}$Sn+$^{124}$Sn collisions at the beam energy of 270 AMeV and an impact parameter $\text{b}=4$ fm, with similar setups as in Ref.~\cite{TMEP:2023ifw}, to illustrate the AMC effect on pion production. We first compare the $2 \leftrightarrow 2$ reaction rates with and without AMC in Fig.~\ref{fig2}, where the Pauli blocking and the $\Delta \leftrightarrow N+\pi$ processes are turned off. As shown in Ref.~\cite{Liu:2023pgc}, the AMC separates the final-state nucleons in elastic NN scatterings and leads to slightly lower densities, and thus lower scattering rates compared to the case without AMC, as shown in the first row of Fig.~\ref{fig2}. In Fig.~\ref{fig2} (c), one sees that this effect also leads to a lower $N+N \rightarrow N+\Delta$ rate in the most compressed stage of heavy-ion collisions, but an enhanced $N+N \rightarrow N+\Delta$ rate is observed in the later expansion stage due to short distances between final-state nucleons after the $N+\Delta \rightarrow N+N$ process as shown in Fig.~\ref{fig1} (d). In the C.M. energy distribution as shown in Fig.~\ref{fig2} (d), the above two effects largely cancel each other. With AMC, the $N+\Delta\rightarrow N+N$ rate is suppressed in the most compressed stage as shown in Fig.~\ref{fig2} (e), due to the larger separation of $N$ and $\Delta$ in the final state of the $N+N\rightarrow N+\Delta$ process as shown in Fig.~\ref{fig1} (c). While there is a small enhancement of $N+\Delta\rightarrow N+N$ rate in the later expansion state, this is due to the larger number of $\Delta$ resonances, and it can be more clearly seen in Fig.~\ref{fig2} (f) that the suppression is for high-mass $\Delta$ resonances while the enhancement is for low-mass $\Delta$ resonances. Due to the dominating suppression effect on the $N+\Delta\rightarrow N+N$ channel with AMC, it can be seen from the bottom row of Fig.~\ref{fig2} that the net $\Delta$ production is enhanced in the early stage and for high-mass $\Delta$ resonances, but only slightly suppressed in the later stage and for low-mass $\Delta$ resonances.

We now take a close look on the isospin dependence of the above effect by monitoring two representative channels $n+n \leftrightarrow p + \Delta^-$ and $p+p \leftrightarrow n+\Delta^{++}$ in Fig.~\ref{fig3}. The $n+n \leftrightarrow p + \Delta^-$ rate is about $2\sim3$ times as the $p+p \leftrightarrow n+\Delta^{++}$ rate, while the suppression and the enhancement effects with AMC are even stronger in the $n+n \leftrightarrow p + \Delta^-$ channel than in the $p+p \leftrightarrow n+\Delta^{++}$ channel in the corresponding time and C.M. energy regions. This is not surprising since we are comparing reaction rates in the dense neutron-rich matter, and one expects that the AMC may affect the pion production in an isospin-dependent way.

The AMC effect on the $\Delta$ decay as well as the inverse process in the same reaction is displayed in Fig.~\ref{fig4}, where the Pauli blocking is still turned off. As discussed before, due to the upper limit of $\Delta$ spin angular momentum, the constraint of AMC suppresses $N+\pi \rightarrow \Delta$ reactions, as shown in the lower panels of Fig.~\ref{fig4}. Given the AMC in $2 \leftrightarrow 2$ channels turned on in this comparison, the suppression of the $N+\pi \rightarrow \Delta$ process leads to fewer $\Delta$ resonances and thus a smaller $\Delta$ decay rate, as shown in the upper panels of Fig.~\ref{fig4}. But this is a secondary effect, and we can see that the constraint of AMC in the $1 \leftrightarrow 2$ channels enhances the net $\Delta$ decay and produces more pions, by comparing the different suppression magnitudes in the upper and lower panels of Fig.~\ref{fig4}.

How multiplicities of $\Delta$ resonances and pions evolve with time in the same reaction is displayed in Fig.~\ref{fig5}, where results with and without AMC for all channels are compared. The constraint of AMC enhances considerably $\Delta$ production, due to the suppression on the $N+\Delta \rightarrow N+N$ channel, and significantly pion production, due to the additional suppression on the $N+\pi \rightarrow \Delta$ channel, as shown Fig.~\ref{fig5} (a). If the Pauli blocking is turned on, the multiplicities for all pion-like particles are reduced, but the AMC effect remains qualitatively similar, as shown Fig.~\ref{fig5} (b).

The final pion multiplicity and $\pi^-/\pi^+$ yield ratio with and without AMC are shown in Fig.~\ref{fig6}, and results for different collision systems are compared. As shown in Fig.~\ref{fig6} (a), the constraint of AMC increases the pion multiplicity by about $75\%$ in Sn+Sn collisions at 270 AMeV, $65\%$ in Au+Au collisions at 400 AMeV, and $38\%$ in Au+Au collisions at 1 AGeV. At higher collision energies, higher densities are reached and more energetic particles are produced, making the $\Delta$ and pion absorption easier and thus less affected by AMC. A larger pion multiplicity generally reduces the $\pi^-/\pi^+$ yield ratio as discussed in Ref.~\cite{TMEP:2023ifw}, and this is also seen in Fig.~\ref{fig6} (b). Suppose the parameters in IBUU in the case without AMC are fitted to reproduce the pion multiplicity data, incorporating the constraint of AMC may overestimate the pion yield, and here we check what if we use a density-dependent cross section for the $N+N \leftrightarrow N+\Delta$ channels as in Ref.~\cite{Song:2015hua}:
\begin{equation}
\sigma_{N+N \leftrightarrow N+\Delta}(\rho) = \sigma_{N+N \leftrightarrow N+\Delta} (0) \exp(-C \rho/\rho_0),
\end{equation}
where the parameter $C$ is adjusted in the case with AMC to reproduce the pion multiplicity in the case without AMC. With this in-medium cross section, it is seen that the $\pi^-/\pi^+$ yield ratio remains almost unchanged in Sn+Sn collisions at 270 AMeV and Au+Au collisions at 400 AMeV, although it is becomes similar to the case without AMC in Au+Au collisions at 1 AGeV. This shows that the constraint of AMC is not just a bulk effect but affects the pion yield in an isospin-dependent way.


To summarize, considering the spin degree of freedom for nucleons and $\Delta$ resonances, we have investigated the AMC effect on pion production in intermediate-energy heavy-ion collisions based on the IBUU transport model. The constraint of AMC suppresses the $N+\Delta \rightarrow N+N$ and $N+\pi \rightarrow \Delta$ channels, and thus enhances considerably $\Delta$ production and significantly pion production. The $\pi^-/\pi^+$ yield ratio is reduced with the constraint of AMC, especially at lower collision energies where the effect can not be simply compensated by a density-dependent $N+N \leftrightarrow N+\Delta$ cross section.


This work is supported by the National Key Research and Development Program of China under Grant No. 2023YFA1606701, the National Natural Science Foundation of China under Grant Nos. 12375125, 11922514, and 11475243, and the Fundamental Research Funds for the Central Universities.

\bibliography{AMC_Delta}

@article{Lattimer:2006xb,
    author = "Lattimer, James M. and Prakash, Maddapa",
    title = "{Neutron Star Observations: Prognosis for Equation of State Constraints}",
    eprint = "astro-ph/0612440",
    archivePrefix = "arXiv",
    doi = "10.1016/j.physrep.2007.02.003",
    journal = "Phys. Rept.",
    volume = "442",
    pages = "109--165",
    year = "2007"
}

@article{Steiner:2004fi,
    author = "Steiner, Andrew W. and Prakash, Madappa and Lattimer, James M. and Ellis, Paul J.",
    title = "{Isospin asymmetry in nuclei and neutron stars}",
    eprint = "nucl-th/0410066",
    archivePrefix = "arXiv",
    reportNumber = "LA-UR-04-6745",
    doi = "10.1016/j.physrep.2005.02.004",
    journal = "Phys. Rept.",
    volume = "411",
    pages = "325--375",
    year = "2005"
}

@article{Baran:2004ih,
    author = "Baran, V. and Colonna, M. and Greco, V. and Di Toro, M.",
    title = "{Reaction dynamics with exotic beams}",
    eprint = "nucl-th/0412060",
    archivePrefix = "arXiv",
    doi = "10.1016/j.physrep.2004.12.004",
    journal = "Phys. Rept.",
    volume = "410",
    pages = "335--466",
    year = "2005"
}

@article{Li:2008gp,
    author = "Li, Bao-An and Chen, Lie-Wen and Ko, Che Ming",
    title = "{Recent Progress and New Challenges in Isospin Physics with Heavy-Ion Reactions}",
    eprint = "0804.3580",
    archivePrefix = "arXiv",
    primaryClass = "nucl-th",
    doi = "10.1016/j.physrep.2008.04.005",
    journal = "Phys. Rept.",
    volume = "464",
    pages = "113--281",
    year = "2008"
}

@article{Oertel:2016bki,
    author = {Oertel, M. and Hempel, M. and Kl{\"a}hn, T. and Typel, S.},
    title = "{Equations of state for supernovae and compact stars}",
    eprint = "1610.03361",
    archivePrefix = "arXiv",
    primaryClass = "astro-ph.HE",
    doi = "10.1103/RevModPhys.89.015007",
    journal = "Rev. Mod. Phys.",
    volume = "89",
    number = "1",
    pages = "015007",
    year = "2017"
}

@article{Li:2002qx,
    author = "Li, Bao-An",
    title = "{Probing the high density behavior of nuclear symmetry energy with high-energy heavy ion collisions}",
    eprint = "nucl-th/0205002",
    archivePrefix = "arXiv",
    doi = "10.1103/PhysRevLett.88.192701",
    journal = "Phys. Rev. Lett.",
    volume = "88",
    pages = "192701",
    year = "2002"
}

@article{Li:2002yda,
    author = "Li, Bao-An",
    title = "{High density behavior of nuclear symmetry energy and high-energy heavy ion collisions}",
    eprint = "nucl-th/0206053",
    archivePrefix = "arXiv",
    doi = "10.1016/S0375-9474(02)01018-7",
    journal = "Nucl. Phys. A",
    volume = "708",
    pages = "365--390",
    year = "2002"
}

@article{Xiao:2008vm,
    author = "Xiao, Zhigang and Li, Bao-An and Chen, Lie-Wen and Yong, Gao-Chan and Zhang, Ming",
    title = "{Circumstantial Evidence for a Soft Nuclear Symmetry Energy at Suprasaturation Densities}",
    eprint = "0808.0186",
    archivePrefix = "arXiv",
    primaryClass = "nucl-th",
    doi = "10.1103/PhysRevLett.102.062502",
    journal = "Phys. Rev. Lett.",
    volume = "102",
    pages = "062502",
    year = "2009"
}

@article{Feng:2009am,
    author = "Feng, Zhao-Qing and Jin, Gen-Ming",
    title = "{Probing high-density behavior of symmetry energy from pion emission in heavy-ion collisions}",
    eprint = "0904.2990",
    archivePrefix = "arXiv",
    primaryClass = "nucl-th",
    doi = "10.1016/j.physletb.2009.12.006",
    journal = "Phys. Lett. B",
    volume = "683",
    pages = "140--144",
    year = "2010"
}

@article{TMEP:2023ifw,
    author = "Xu, Jun and others",
    collaboration = "TMEP",
    title = "{Comparing pion production in transport simulations of heavy-ion collisions at 270AMeV under controlled conditions}",
    eprint = "2308.05347",
    archivePrefix = "arXiv",
    primaryClass = "nucl-th",
    doi = "10.1103/PhysRevC.109.044609",
    journal = "Phys. Rev. C",
    volume = "109",
    number = "4",
    pages = "044609",
    year = "2024"
}

@article{Xie:2013np,
    author = "Xie, Wen-Jie and Su, Jun and Zhu, Long and Zhang, Feng-Shou",
    title = "{Symmetry energy and pion production in the Boltzmann-Langevin approach}",
    doi = "10.1016/j.physletb.2012.12.021",
    journal = "Phys. Lett. B",
    volume = "718",
    pages = "1510--1514",
    year = "2013"
}

@article{Song:2015hua,
    author = "Song, Taesoo and Ko, Che Ming",
    title = "{Modifications of the pion-production threshold in the nuclear medium in heavy ion collisions and the nuclear symmetry energy}",
    doi = "10.1103/PhysRevC.91.014901",
    journal = "Phys. Rev. C",
    volume = "91",
    number = "1",
    pages = "014901",
    year = "2015"
}

@article{Cozma:2014yna,
    author = "Cozma, M. D.",
    title = "{The impact of energy conservation in transport models on the $\pi^-/\pi^+$ multiplicity ratio in heavy-ion collisions and the symmetry energy}",
    eprint = "1409.3110",
    archivePrefix = "arXiv",
    primaryClass = "nucl-th",
    doi = "10.1016/j.physletb.2015.12.015",
    journal = "Phys. Lett. B",
    volume = "753",
    pages = "166--172",
    year = "2016"
}

@article{Ikeno:2016xpr,
    author = "Ikeno, Natsumi and Ono, Akira and Nara, Yasushi and Ohnishi, Akira",
    title = "{Probing neutron-proton dynamics by pions}",
    eprint = "1601.07636",
    archivePrefix = "arXiv",
    primaryClass = "nucl-th",
    reportNumber = "YITP-16-6",
    doi = "10.1103/PhysRevC.93.044612",
    journal = "Phys. Rev. C",
    volume = "93",
    number = "4",
    pages = "044612",
    year = "2016",
    note = "[Erratum: Phys.Rev.C 97, 069902 (2018)]"
}

@article{Zhang:2017mps,
    author = "Zhang, Zhen and Ko, Che Ming",
    title = "{Medium effects on pion production in heavy ion collisions}",
    eprint = "1701.06682",
    archivePrefix = "arXiv",
    primaryClass = "nucl-th",
    doi = "10.1103/PhysRevC.95.064604",
    journal = "Phys. Rev. C",
    volume = "95",
    number = "6",
    pages = "064604",
    year = "2017"
}

@article{Hong:2013yva,
    author = "Hong, Jun and Danielewicz, P.",
    title = "{Subthreshold pion production within a transport description of central Au + Au collisions}",
    eprint = "1307.7654",
    archivePrefix = "arXiv",
    primaryClass = "nucl-th",
    doi = "10.1103/PhysRevC.90.024605",
    journal = "Phys. Rev. C",
    volume = "90",
    number = "2",
    pages = "024605",
    year = "2014"
}

@article{Xu:2013aza,
    author = "Xu, Jun and Chen, Lie-Wen and Ko, Che Ming and Li, Bao-An and Ma, Yu-Gang",
    title = "{Energy dependence of pion in-medium effects on the {\ensuremath{\pi}}{\ensuremath{-}}/{\ensuremath{\pi}}+ ratio in heavy-ion collisions}",
    eprint = "1305.0091",
    archivePrefix = "arXiv",
    primaryClass = "nucl-th",
    doi = "10.1103/PhysRevC.87.067601",
    journal = "Phys. Rev. C",
    volume = "87",
    number = "6",
    pages = "067601",
    year = "2013"
}

@article{Xu:2009fj,
    author = "Xu, Jun and Ko, Che Ming and Oh, Yongseok",
    title = "{Isospin-dependent pion in-medium effects on charged pion ratio in heavy ion collisions}",
    eprint = "0906.1602",
    archivePrefix = "arXiv",
    primaryClass = "nucl-th",
    doi = "10.1103/PhysRevC.81.024910",
    journal = "Phys. Rev. C",
    volume = "81",
    pages = "024910",
    year = "2010"
}

@article{Gale:1990zz,
    author = "Gale, C. and Das Gupta, S.",
    title = "{Conservation laws and nuclear transport models}",
    doi = "10.1103/PhysRevC.42.1577",
    journal = "Phys. Rev. C",
    volume = "42",
    pages = "1577",
    year = "1990"
}

@article{Liu:2023pgc,
    author = "Liu, Rong-Jun and Xu, Jun",
    title = "{Revisiting Angular Momentum Conservation in Transport Simulations of Intermediate-Energy Heavy-Ion Collisions}",
    eprint = "2301.07907",
    archivePrefix = "arXiv",
    primaryClass = "nucl-th",
    doi = "10.3390/universe9010036",
    journal = "Universe",
    volume = "9",
    number = "1",
    pages = "36",
    year = "2023"
}

@article{Liu:2023nkm,
    author = "Liu, Rong-Jun and Xu, Jun",
    title = "{Spin dynamics in intermediate-energy heavy-ion collisions with rigorous angular momentum conservation}",
    eprint = "2311.13769",
    archivePrefix = "arXiv",
    primaryClass = "nucl-th",
    doi = "10.1103/PhysRevC.109.014615",
    journal = "Phys. Rev. C",
    volume = "109",
    number = "1",
    pages = "014615",
    year = "2024"
}

@article{Jacob:1959at,
    author = "Jacob, M. and Wick, G. C.",
    title = "{On the General Theory of Collisions for Particles with Spin}",
    doi = "10.1006/aphy.2000.6022",
    journal = "Annals Phys.",
    volume = "7",
    pages = "404--428",
    year = "1959"
}

@book{wigner2012group,
  title={Group theory: and its application to the quantum mechanics of atomic spectra},
  author={Wigner, Eugene},
  year={1959},
}

@article{Bertsch:1988ik,
    author = "Bertsch, G. F. and Das Gupta, S.",
    title = "{A Guide to microscopic models for intermediate-energy heavy ion collisions}",
    doi = "10.1016/0370-1573(88)90170-6",
    journal = "Phys. Rept.",
    volume = "160",
    pages = "189--233",
    year = "1988"
}

@article{Lenk:1989zz,
    author = "Lenk, R. J. and Pandharipande, V. R.",
    title = "{Nuclear mean field dynamics in the lattice Hamiltonian Vlasov method}",
    doi = "10.1103/PhysRevC.39.2242",
    journal = "Phys. Rev. C",
    volume = "39",
    pages = "2242--2249",
    year = "1989"
}
\end{document}